 \documentstyle[aps,prl,epsf,twocolumn]{revtex}
\begin{document}
\draft
\title{
%======================================================================%
  Complete Numerical Solution of the Temkin-Poet Three-Body Problem
%======================================================================%
}
\author{
%======================================================================%
                 S. Jones and A. T. Stelbovics
%======================================================================%
}
\address{
%======================================================================%
     Centre for Atomic, Molecular and Surface Physics,
%    Physics and Engineering Studies,
     Division of Science,
     Murdoch University, Perth 6150, Australia
%======================================================================%
}
\date{\today}
\maketitle

%======================================================================%
\begin{abstract}
%======================================================================%
Although the convergent close-coupling (CCC) method has achieved
unprecedented success in obtaining accurate theoretical cross sections
for electron-atom scattering, it generally fails to yield converged
energy distributions for ionization.
Here we report converged energy distributions for ionization of
$\rm{H}(1s)$ by numerically integrating Schr\"{o}dinger's equation
subject to correct asymptotic boundary conditions for the Temkin-Poet
model collision problem, which neglects angular momentum.
Moreover, since the present method is complete, we obtained convergence
for all transitions in a single calculation (excluding the very highest
Rydberg transitions, which require integrating to infinitely-large
distances; these cross sections may be accurately obtained from
lower-level Rydberg cross sections using the $1/n^3$ scaling law).
Complete results, accurate to 1\%, are presented for impact energies of
54.4 and 40.8 eV, where CCC results are available for comparison.
\end{abstract}

%======================================================================%
%======================================================================%

\narrowtext

%======================================================================%
%  INTRODUCTION
%======================================================================%
The Temkin-Poet (TP) model \cite{Temkin,Poet78} of electron-hydrogen
scattering is now widely regarded as an ideal testing ground for the
development of general methods intended for the full three-body Coulomb
problem.
Although only $s$-states are included for both projectile and atomic
electrons, this model problem still contains most of the features that
make the real scattering problem hard to solve.
Indeed, even in this simplified model, converged energy distributions
for ionization can not generally be obtained via the close-coupling
formalism \cite{CCC}.
Any general method that can not obtain complete, converged results for
this model problem will face similar difficulties when applied to the
full electron-hydrogen system.
Therefore we believe it is essential to develop a numerical method
capable of solving the TP model completely before angular momentum is
included.
Here we report such a method.
Complete, precision results for $e^- + {\rm H(1s)}$, accurate to 1\%,
are presented for total energies of 3 and 2 Rydbergs (Ryd).
Atomic units (Ryd energy units) are used throughout this work unless
stated otherwise.

%======================================================================%
%  THEORY
%======================================================================%
Our numerical method may be summerized as follows.
The model Schr\"{o}dinger equation is integrated outwards from the
atomic center on a grid of fixed spacing $h$.
The number of difference equations is reduced each step outwards using
an algorithm due to Poet \cite{Poet80}, resulting in a propagating
solution of the partial-differential equation.
By imposing correct asymptotic boundary conditions on this general,
propagating solution, the particular solution that physically
corresponds to scattering is obtained along with the scattering
amplitudes.

The Schr\"{o}dinger equation in the TP model is given by
\begin{equation}
   \left(
   \frac{\partial^2}{\partial x^2}
  +\frac{\partial^2}{\partial y^2}
  +\frac{2}{\min(x,y)}
  +E
  \right)
  \Psi(x,y) = 0,
\label{pde}
\end{equation}
with boundary conditions
\begin{equation}
   \Psi(x,0) =
   \Psi(0,y) = 0
\end{equation}
and symmetry condition
\begin{equation}
   \Psi(y,x) = \pm\Psi(x,y),
\label{sym}
\end{equation}
depending on whether the two electrons form a singlet ($+$) or triplet
($-$) spin state.
Eq. (\ref{pde}) is separable in the two regions $x \ge y$ and
$x \le y$.
Because of the symmetry condition (\ref{sym}), we can solve
Eq. (\ref{pde}) in just one of these regions and this is sufficient to
determine all of the scattering information.
For brevity, we do not explicitly indicate the total spin since the
singlet and triplet cases require completely separate calculations.
For $x \ge y$, the wave function may be written
\begin{eqnarray}
   \Psi(x,y) =
             \psi_{\epsilon_i}(y)
             e^{-i k_{\epsilon_i} x}
            +\sum_{j=1}^{\infty} C_{\epsilon_j i}
             \psi_{\epsilon_j}(y)
             e^{i k_{\epsilon_j} x} \nonumber \\
            +\int_0^{\infty} d\epsilon_b C_{\epsilon_b i}
             \psi_{\epsilon_b}(y)
             e^{i k_{\epsilon_b} x}.
\label{wf_form}
\end{eqnarray}
The $\psi_{\epsilon}$ are bound and continuum states of the
hydrogen atom with zero angular momentum:
\begin{eqnarray}
   \psi_{\epsilon}(y) = y e^{-qy} {_1F_1}(1-1/q,2;2qy).
\end{eqnarray}
Here $q^2 = -\epsilon$, where $\epsilon$ is the inner electron energy,
and ${_1F_1}$ is the confluent hypergeometric function.
The momenta in (\ref{wf_form}) are fixed by energy conservation
according to
\begin{equation}
       \epsilon_i + k_{\epsilon_i}^2
     = \epsilon_j + k_{\epsilon_j}^2
     = \epsilon_b + k_{\epsilon_b}^2
     = E,
\end{equation}
where $E>0$ is the total energy.
The $C_{\epsilon i}$ are related to S-matrix elements by
normalization factors: 
\begin{equation}
   {\rm S}_{\epsilon_j i} = -\left(\frac{k_{\epsilon_j}}
                                        {k_{\epsilon_i}}\right)^{1/2}
                             \left(\frac{j}{i}\right)^{3/2}
                             C_{\epsilon_j i}
\end{equation}
for discrete transitions and
\begin{equation}
   {\rm S}_{\epsilon_b i} = -\left(\frac{k_{\epsilon_b}}
                                        {k_{\epsilon_i}}\right)^{1/2}
                             \left(\frac{1}{i}\right)^{3/2}
                             \left[\frac{1-e^{-2\pi/k}}{4k}\right]^{1/2}
                             C_{\epsilon_b i}
\end{equation}
for ionization, where $k = \sqrt{\epsilon_b}$.
Cross sections are obtained from S-matrix elements in the usual manner.

To convert the partial-differential equation (\ref{pde}) into 
difference equations we impose a grid of fixed spacing $h$
and approximate derivatives by finite differences.
After applying the Numerov scheme in both the $x$ and $y$ directions,
our difference equations have the form \cite{Poet80}
\begin{equation}
   {\bf A}^{(i)} {\bf \cdot} \overline{\Psi}^{(i-1)}+
   {\bf B}^{(i)} {\bf \cdot} \overline{\Psi}^{(i)}+
   {\bf C}^{(i)} {\bf \cdot} \overline{\Psi}^{(i+1)}={\bf 0},
\label{form}
\end{equation}
Here we have collected the various $\Psi^{(i)}_j, j = 1,2,\dots,i$,
where $\Psi^{(i)}_j \equiv \Psi(x=ih,y=jh)$,
into a vector $\overline{\Psi}^{(i)}$.
The matrices ${\bf A}^{(i)}$, ${\bf B}^{(i)}$ and
${\bf C}^{(i)}$ are completely determined by the formulas given by
Poet \cite{Poet80}.

At each value of $i$ we can solve our equations if we apply symbolic
boundary conditions at $i+1$ [solve for $\Psi^{(i)}_j$ in terms of
$\Psi^{(i+1)}_j$ ($j=1,2,\dots,i$)].
This procedure yields a propagation matrix ${\bf D}^{(i)}$:
\begin{equation}
   \overline\Psi^{(i)} = {\bf D}^{(i)} \cdot \overline\Psi^{(i+1)}.
\label{D}
\end{equation}
We can obtain a recursion relation for ${\bf D}^{(i)}$ by using
(\ref{D}) to eliminate $\overline\Psi^{(i-1)}$ from equation
(\ref{form}):
\begin{equation}
   \left[ {\bf B}^{(i)} + {\bf A}^{(i)} \cdot {\bf D}^{(i-1)} \right]
          \cdot \overline\Psi^{(i)} =
         -{\bf C} ^{(i)} \cdot \overline\Psi^{(i+1)}.
\label{diff2}
\end{equation}
Comparing (\ref{diff2}) with (\ref{D}),
\begin{equation}
   {\bf D}^{(i)}=
   -\left[{\bf B}^{(i)}+{\bf A}^{(i)} \cdot {\bf D}^{(i-1)} \right]^{-1}
   \cdot {\bf C}^{(i)}.
\end{equation}
Thus each ${\bf D}^{(i)}$ is determined from the previous one
(${\bf D}^{(1)}$ can be determined by inspection).

In the asymptotic region, the form of the wave function is known and is
given in terms of the
            $C_{\epsilon i}$ by
\begin{eqnarray}
 &&\overline\Psi^{(i)} \sim
   {\bf I}^{(i)} + {\bf R}^{(i)} \cdot {\bf C}.
\label{bc}
\end{eqnarray}
Here the matrix ${\bf I}^{(i)}$ contains the incident part of the
asymptotic solution while ${\bf R}^{(i)}$ contains the reflected part.
The asymptotic solution is identical to the full solution,
Eq. (\ref{wf_form}), except that the quadrature over the continuum
extends only up to the total energy $E$.
The infinite summation over discrete channels is truncated to some
finite integer $N_d$ and the quadrature over the two-electron continuum
is performed prior to matching by first writing the $C_{\epsilon_b i}$
as a power series in $\epsilon_b$:
\begin{equation}
            C_{\epsilon_b i} \approx
                \sum_{n=1}^{N_c}
                {\rm c}_{n i}
                \epsilon_b^n.
\end{equation}
The matching procedure then determines the (in practice, much smaller
set of) coefficients ${\rm c}_{n i}$, rather than the $C_{\epsilon_b i}$
directly, which eliminates ill conditioning \cite{Poet80}.

To extract an $N \times N$ coefficient matrix, where $N=N_d+N_c$, we
need only $N$ of the $i$ equations (\ref{D}).
Alternatively, one may use all $i$ equations as in Poet \cite{Poet80}.
In this case, the system of equations is overdetermined.
Nevertheless, a solution can be found by the standard method of
minimizing the sum of the squares of the residuals [the differences
between the left- and right-hand sides of equations (\ref{D})].
Previously \cite{JS}, we found that the least-squares method is generally
stabler than keeping any subset of just $N$ equations (\ref{D}).

%======================================================================%
%  NUMERICAL METHOD
%======================================================================%
Our numerical method is stable and rapidly convergent.
For a given grid spacing $h$, we established convergence in propagation
distance by performing the matching every 40 a.u. until convergence was
obtained.
At each matching radius, both the number of discrete channels $N_d$ and
the number of expansion functions for the continuum $N_c$ were varied
to obtain convergence.
Finally, the entire calculation was repeated for a finer grid
(using one-half the original grid spacing $h$).

The biggest advantage of having a general, propagating solution is that
once the grid spacing is chosen, a ``single'' calculation is all that
is needed to establish convergence for the remaining numerical
parameters.
This is because the D-matrix, the calculation of which consumes nearly
all the computational effort, is independent of asymptotic boundary
conditions.
Thus, in a typical calculation, the same D-matrix is used for,
{\em e.g.}, $N_c=0,1,\dots,9$ while $N_d$ runs from 1 to 30.
This would have required 300 completely separate calculations (each
taking about the same time as our ``one'' calculation) had we solved the
original global matrix equations (\ref{form}).

%======================================================================%
%  RESULTS
%======================================================================%
We have performed complete calculations for electrons colliding with
atomic hydrogen at impact energies of 54.4 and 40.8 eV
(total energies of 3 and 2 Ryd, respectively).
In Table I, we present our calculated cross sections for
$e^-+{\rm H}(1s) \rightarrow e^-+{\rm H}(ns), n \le 8$.
The grid spacing is $h = 1/5$ a.u. (results using one-half this spacing
differed by less than 0.1\% for discrete excitations and 0.5\% for
elastic scattering).
One of the advantages of our direct approach is that we are able to
obtain the amplitudes for higher-level (Rydberg) transitions as easily
as those for low-level excitations, provided the matching radius is
large enough to enclose the final Rydberg state.
This is in contrast to some other approaches, such as the CCC,
which lose accuracy for higher-level transitions.

%======================================================================%
\begin{table}
%======================================================================%
\caption{
Cross sections ($\pi a_o^2$; superscripts indicate powers of 10) for
$e^-+{\rm H}(1s) \rightarrow e^-+{\rm H}(ns)$, for impact energies of
54.4 and 40.8 eV.
}
\begin{tabular}{ccccc}
   & \multicolumn{2}{c}{54.4 eV} & \multicolumn{2}{c}{40.8 eV}\\
n  & Singlet & Triplet & Singlet & Triplet \\ \hline

1&
6.47$^{-2}$&
4.07$^{-1}$&
8.58$^{-2}$&
6.34$^{-1}$ \\

2&
4.66$^{-3}$&
4.04$^{-3}$&
8.09$^{-3}$&
5.08$^{-3}$ \\

3&
1.22$^{-3}$&
8.39$^{-4}$&
2.15$^{-3}$&
9.88$^{-4}$ \\

4&
4.92$^{-4}$&
3.13$^{-4}$&
8.74$^{-4}$&
3.59$^{-4}$ \\

5&
2.48$^{-4}$&
1.52$^{-4}$&
4.41$^{-4}$&
1.71$^{-4}$ \\

6&
1.42$^{-4}$&
8.52$^{-5}$&
2.53$^{-4}$&
9.55$^{-5}$ \\

7&
8.89$^{-5}$&
5.27$^{-5}$&
1.58$^{-4}$&
5.88$^{-5}$ \\

8&
5.94$^{-5}$&
3.49$^{-5}$&
1.06$^{-4}$&
3.88$^{-5}$

\end{tabular}
\label{table1}
\end{table}

%======================================================================%
\begin{figure}
%======================================================================%
\epsfxsize=7.5truecm
\epsffile{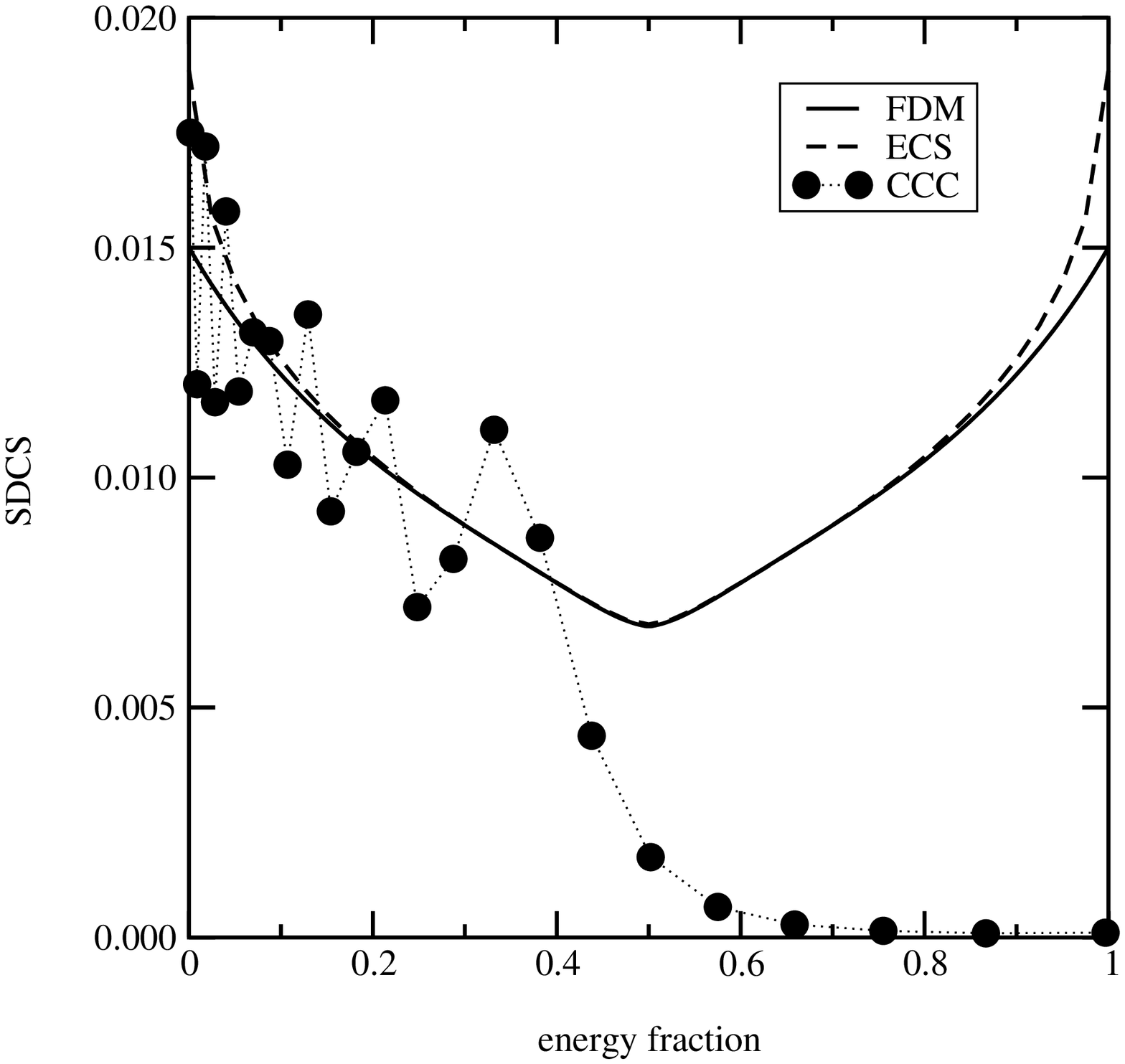}
\caption{
Singlet SDCS ($\pi a_o^2/{\rm Ryd}$) vs. the energy fraction
$\epsilon_b/E$ for an impact energy of 54.4 eV.
The total ionization cross sections from FDM, ECS, and CCC are
$1.50^{-2}$,$1.54^{-2}$, and $1.48^{-2}$ ($\pi a_o^2$), respectively.
}
\label{fig1}
\end{figure}

%======================================================================%
\begin{figure}
%======================================================================%
\epsfxsize=7.5truecm
\epsffile{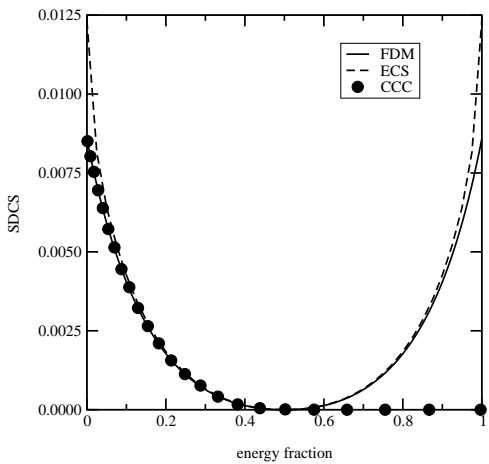}
\caption{
Same as Fig. 1 for the triplet case.
The total ionization cross sections from FDM, ECS, and CCC are
$3.11^{-3}$, $3.39^{-3}$, and $3.21^{-3}$ ($\pi a_o^2$), respectively.
}
\label{fig2}
\end{figure}

%======================================================================%
\begin{figure}
%======================================================================%
\epsfxsize=7.5truecm
\epsffile{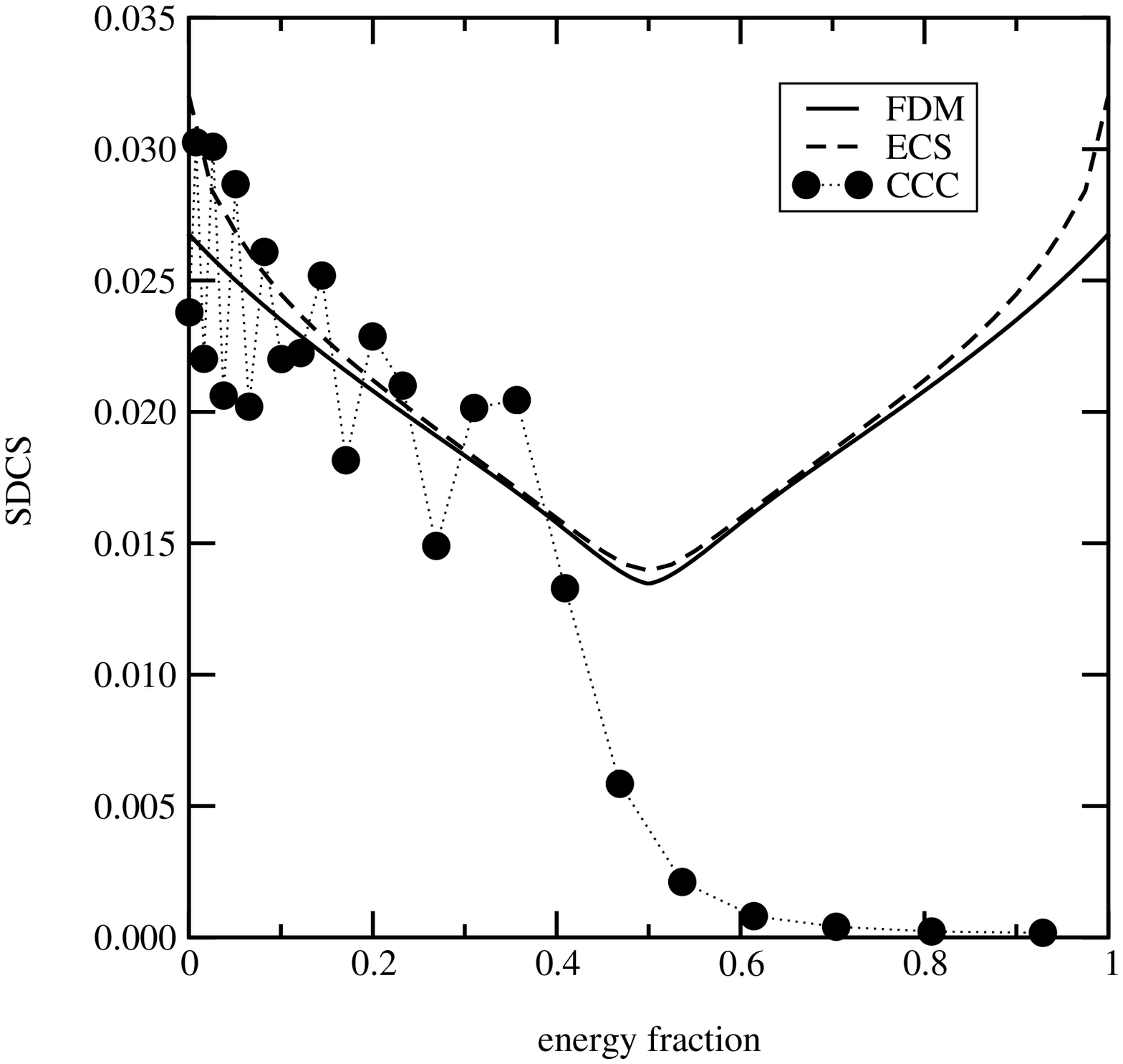}
\caption{
Singlet SDCS ($\pi a_o^2/{\rm Ryd}$) vs. the energy fraction
$\epsilon_b/E$ for an impact energy of 40.8 eV.
The total ionization cross sections from FDM, ECS, and CCC are
$1.97^{-2}$, $2.04^{-2}$, and $2.02^{-2}$ ($\pi a_o^2$), respectively.
}
\label{fig3}
\end{figure}

%======================================================================%
\begin{figure}
%======================================================================%
\epsfxsize=7.5truecm
\epsffile{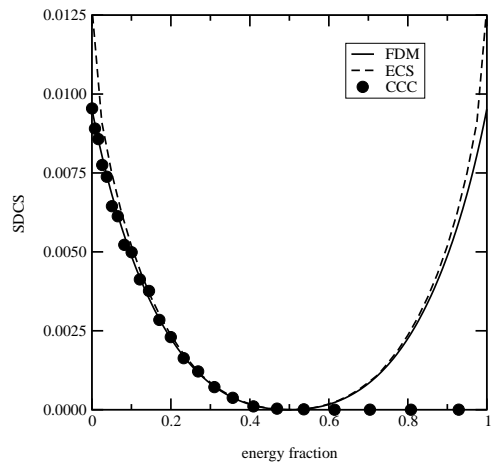}
\caption{
Same as Fig. 3 for the triplet case.
The total ionization cross sections from FDM, ECS, and CCC are
$2.47^{-3}$, $2.70^{-3}$, and $2.50^{-3}$ ($\pi a_o^2$), respectively.
}
\label{fig4}
\end{figure}

In Figures 1-4, we present our results (labeled FDM for
finite-difference method) for the single-differential cross section
(SDCS).
For a total energy of 3 Ryd, 240 a.u. proved to be a sufficient matching
radius to get convergence of the SDCS and for E = 2 Ryd, a radius of 360
a.u. was required.
The SDCS is more sensitive to the number of expansion functions for the
continuum than the other observables, particularly about
$\epsilon_b=E/2$.
Nevertheless, convergence to better than 1\% was readily obtained using
7-8 functions
(the largest discrepancy in the SDCS between $N_c=7$ and $N_c=8$ was
smaller than 0.3\%; even using just 6 expansion function gave results
accurate to 1\%).

Also shown in Figs. 1-4 are the results of convergent close-coupling
(CCC) calculations \cite{CCC}.
The CCC method of Bray \cite{CCC} employs a ``distinguishable electron''
prescription, which produces energy distributions that are not symmetric
about $\epsilon_b = E/2$.
Stelbovics \cite{Stelbovics} has shown that a properly symmetrized CCC
amplitude yields SDCS that {\em are} symmetric about $E/2$ as well as
being four times larger at $\epsilon_b = E/2$ than those assuming
distinguishable electrons.
(Note that our singlet FDM results at $\epsilon_b = E/2$ are about four
times larger than the corresponding CCC results.)
Other than making the energy distributions symmetric, it is clear from
the figures that symmetrization (coherent summation of the CCC
amplitudes $C_{\epsilon_b i}$ and $C_{\epsilon_a i}$,
where $\epsilon_a = E - \epsilon_b$, which correspond
to physically indistinguishable processes) will significantly affect
only singlet scattering (and then only near $\epsilon_b = E/2$), since
$C_{\epsilon_b i}$ is practically zero for $\epsilon_b > E/2$.
For singlet scattering, the CCC oscillates about the true value of SDCS,
except near (and beyond) $\epsilon_b = E/2$.
CCC results for triplet scattering, on the other hand, are in very good
agreement with our results for $0 \le \epsilon_b \le E/2$.

Some very recent results from Baertschy {\em et al.} \cite{ECS} have
also been included in the figures.
Baertschy {\em et al.} rearrange the Schr\"{o}dinger equation to solve
for the outgoing scattered wave.
They use a two-dimensional grid like ours, but scale the coordinates by
a complex phase factor beyond a certain radius where the tail of the
Coulomb potential is ignored.
As a result, the scattered wave decays like an ordinary bound state
beyond this cut-off radius, which makes the asymptotic boundary
conditions very simple.
By computing the outgoing flux directly from the scattered wave at
several large cut-off radii, and extrapolating to infinity, they obtain
the single-differential ionization cross section without having to use
Coulomb three-body boundary conditions.
This method, called exterior complex scaling (ECS), has just been
extended to the full electron-hydrogen ionization problem \cite{BRM}.
It is seen from Figs. 1-4 that the ECS results are in good agreement
with our FDM results except when the energy fraction $\epsilon_b/E$
approaches 0 or 1.
Baertschy {\em et al.} \cite{ECS} note that their method may be
unreliable as $\epsilon_b$ approaches 0 or $E$ due to ``contamination''
of the ionization flux by contributions from discrete excitations.

We note also the recent work of Miyashita {\em et al.} \cite{MKW}, who
have presented SDCS for total energies of 4, 2, and 0.1 Ryd using two
different methods.
One produces an asymmetric energy distribution similar to that of CCC
while the other gives a symmetric distribution.
Both contain oscillations.
The mean of their symmetric curve at $E=2$ Ryd (40.8 eV impact energy)
is in reasonable agreement with our calculations.

%======================================================================%
%  CONCLUSION
%======================================================================%
In conclusion, we have presented complete, precision results for the
Temkin-Poet electron-hydrogen scattering problem for impact energies of
54.4 and 40.8 eV.
It may be possible to improve the speed of the present method by using
a variable-spaced grid, like that used by Botero and Shertzer \cite{BS}
in their finite-element analysis (this would greatly reduce storage
requirements as well).
Once we have optimized our code for this simplified model we will
proceed to include angular momentum.
When angular momentum is included, the ionization boundary condition
is no longer separable and this is the major challenge for generalizing
the present approach to the full electron-hydrogen scattering problem.

%======================================================================%
%  ACKNOWLEDGEMENTS
%======================================================================%
\thanks{
The authors gratefully acknowledge the financial support of the
Australian Research Council for this work.
}

%======================================================================%

\end{document}